\documentclass[prd]{revtex4}
\usepackage{graphics}
\usepackage{epsfig}
\usepackage[T2A]{fontenc}

\newcommand\ba{\begin{eqnarray}}
\newcommand\ea{\end{eqnarray}}
\newcommand\nn{\nonumber}

\newcommand{\br}[1]{\left( #1 \right)}
\newcommand{\brs}[1]{\left[ #1 \right]}

\newcommand{\brm}[1]{\left| #1 \right|}
\newcommand{\TeV}{~\mbox{TeV}}

\newcommand{\MeV}{~\mbox{MeV}}

\newcommand{\eV} {~\mbox{eV}}
\newcommand{\mb} {~\mbox{mb}}
\newcommand{\nb} {~\mbox{nb}}
\newcommand{\pb} {~\mbox{pb}}

\newcommand{\ab} {~\mbox{ab}}

\begin{document}

\Large

\title{QED peripheral mechanism of pair production at colliders}

\author{A.~I.~Ahmadov}
\email{Azad.Ahmedov@sunse.jinr.ru}
\affiliation{Joint Institute for Nuclear Research, Dubna, Russia}
\affiliation{Institute of Physics, Azerbaijan National Academy of Sciences, Baku, Azerbaijan}

\author{M.~V.~Galynskii}
\email{galynski@dragon.bas-net.by}
\affiliation{Institute of Physics, Minsk, Belorussia}

\author{Yu.~M.~Bystritskiy}
\email{bystr@theor.jinr.ru}
\affiliation{Joint Institute for Nuclear Research, Dubna, Russia}

\author{E.~A.~Kuraev}
\email{kuraev@theor.jinr.ru}
\affiliation{Joint Institute for Nuclear Research, Dubna, Russia}

\author{M.~G.~Shatnev}
\email{mshatnev@yahoo.com}
\affiliation{Kharkov Institute of Phys. and Tech., Kharkov, 61108, Ukraine}

\begin{abstract}
Cross section of the processes of neutral pion production as well as pairs of
charged fermions and bosons in peripherical
interaction of leptons, photons are calculated in main logarithmical approximation.
We investigate the phase volumes and differential cross sections. The differential cross section
of few neutral pions and a few pairs production are written down explicitly.
Considering the academic problem of summation on the number of pairs for the case of massless
particles we reproduce the known results obtained in seventies of last century.
The possibility to construct the generator for Monte-Carlo modeling of this processes
basing of this results are discussed.
\end{abstract}

\maketitle

\section{Introduction}

The motivation of this paper is the need of creation of realistic
Monte-Carlo generators describing the multiparticle production at
high energy colliders.
We consider the case of peripherical interaction of initial particles
which corresponds to the kinematics of final particles which group
a set of jets flying in the wide range of rapidities --
starting from fragmentation region up to pionization region.

Keeping in mind the possibility to construct the $\gamma-e$ and
$\gamma-\gamma$ colliders on the basis of the $e^+e^-$-colliders
using the backwards scattering of laser beam we consider the processes
which may be relevant for experiments on this facilities.

Formulae, obtained below can be applied to the case of $\gamma p$, $ep$ and $p(\bar{p})p$ colliders
with a straightforward generalization.

Processes of charged pairs production at high energy beams
$a-b$ collisions where $a,b=\gamma,e^\pm$
have a set of general properties. To remind them let consider process of $e^+e^-$ pair production
at high energy $e^\pm e^-$ collisions in peripheral kinematics.
Matrix element of production of a pair of charged particles in high energy electron-positron collisions
(center of mass of initial colliding beams is implied)
\ba
e^+(p)+e^-(k)\to e^+(p')+e^-(k')+q(q_-)+\bar q(q_+),
\ea
\ba
s=2pk\gg -q_1^2=-(k-k')^2 \sim -q_2^2=(p-p')^2
\sim q_\pm^2 = m^2,
\ea
\ba
k^2=m_1^2, \qquad p^2=m_2^2,
\ea
have a form (here and further we consider the so called
<<two-photon mechanism>> of pair production) (see Fig.~\ref{Fig1}):
\begin{figure}
\includegraphics{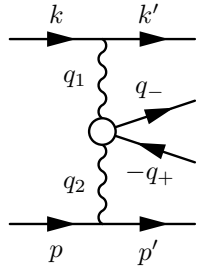}
\caption{The two-photon mechanism of production of a pion pair.}
\label{Fig1}
\end{figure}

\ba
M^{i}=\frac{(4\pi\alpha)^2}{q_1^2 q_2^2}
\brs{\bar u(p') \gamma^\mu u(p)}\brs{\bar v(k') \gamma^\nu v(k)}
O^{i}_{\mu_1\nu_1}g_{\mu\mu_1}g_{\nu\nu_1},
\label{InitialMatrixElement}
\qquad i=e,\pi.
\ea
We use Gribov representation for photon Green function
\ba
q_{\mu\nu}=q_{\mu\nu}^\perp+
\frac{1}{s}\br{p_\mu k_\nu+p_\nu k_\mu}
\label{GribovRepresentation}
\ea
and Sudakov decomposition of 4-momenta
\ba
&&
q_1=\beta_1 \tilde{k}+q_1^\perp, \qquad
q_2=\alpha_2 \tilde{p}+q_2^\perp, \qquad \\
&&q_\pm=\alpha_\pm \tilde{p}+\beta_\pm \tilde{k}+q^\perp_\pm, \qquad
s \alpha_\pm \beta_\pm = \vec{q}_\pm^2 + m^2. \nn
\ea
over <<almost light-cone>>
4-vectors $\tilde{p}=p-k(m_2^2/s)$, $\tilde{k}=k-p(m_1^2/s)$.
Further the tilde subscripts will be omitted.

Keeping in mind the conservation law,
$q_1+q_2=q_++q_-$ and the hierarchy of Sudakov parameters,
$\beta_1 \gg \beta_2$, $\alpha_1 \ll \alpha_2$, the
matrix element (\ref{InitialMatrixElement}) can be written down in the form
\ba
M^{i}=s\frac{(8\pi\alpha)^2}{q^2_1q^2_2}N_1 N_2 \Phi^i(q_1,q_2),\qquad i=e,\pi
\ea
\ba
N_1=\frac{1}{s}\bar{v}(k')\hat{p}v(k), \qquad
N_2=\frac{1}{s}\bar{u}(p')\hat{k}u(p),\label{NeDefinition}
\ea
\ba
\sum|N_1|^2=\sum|N_2|^2=2,
\ea
where factor $\Phi^i(q_1,q_2)=\Phi^i_{12}=(1/s)k^\mu p^\nu O_{\mu\nu}^i$
is the light-cone projection of amplitudes of the subprocess of
charged particles pair production from two virtual photons:
$\gamma^*(q_1)\gamma^*(q_2) \to q(q_-)\bar q(q_+)$.
It is important to note that the quantity $\Phi^i_{12}$
is finite in limit $s \to \infty$ as well as $N_{1,2}$.

\section{The subprocess $\gamma^*(q_1)\gamma^*(q_2) \to q(q_-)\bar q(q_+)$}

In case of electron-positron pair production we have:
\ba
\Phi^e_{12} =
\frac{1}{s} \bar u(q_-)\br{
\hat k \frac{\hat q_- - \hat q_1 + m}{-D_1/x_+} \hat p
+
\hat p \frac{-\hat q_+ + \hat q_1 + m}{-D_2/x_-} \hat k
}v(q_+),
\label{InitialPhiE}
\ea
where
\ba
D_1&=&-x_+\br{\br{q_--q_1}^2-m^2}=\vec{q}_+^2-2x_+\vec{q}_+\vec{q}_2+x_+\vec{q}_2^2 + m^2, \nn\\
D_2&=&-x_-\br{\br{q_--q_2}^2-m^2}=\vec{q}_-^2-2x_-\vec{q}_-\vec{q}_2+x_-\vec{q}_2^2 + m^2, \nn
\ea
where $x_{\pm}$
are the energy fractions of pair components
($x_{\pm}=\alpha_\pm/\alpha_2$, $x_++x_-=1$) and $\vec{q}_i$ is two-dimensional Euclidean vector.
Using Dirac equation $\hat q_- u(q_-) = m u(q_-)$, $\hat q_+ v(q_+) = -m v(q_+)$
we can rewrite formula (\ref{InitialPhiE}) as:
\ba
\Phi^e_{12} =
\frac{1}{s} \bar u(q_-)\br{
A \hat p + B \hat k \hat q_1 \hat p + C \hat p \hat q_1 \hat k
}v(q_+),
\label{PhiE}
\ea
where
\ba
A=sx_+x_-\br{\frac{1}{D_2}-\frac{1}{D_1}}, \qquad
B=\frac{x_+}{D_1}, \qquad
C=-\frac{x_-}{D_2}.\nn
\ea
In case of pion pair production performing the same algebra we obtain:
\ba
\Phi^\pi_{12}=x_+x_-\br{\frac{-2\vec{q}_+\vec{q}_2+\vec{q}_2^2}{D_1}+
\frac{-2\vec{q}_-\vec{q}_2+\vec{q}_2^2}{D_2}}.
\label{InitialPhiPi}
\ea
It can be checked that both $\Phi^e_{12}$, $\Phi^\pi_{12}$
tends to zero at $\vec q_1 \to 0$ or $\vec q_2 \to 0$.
Below we will use the impact factor for production of
charged particles pair with invariant mass $s_1=(q_++q_-)^2$:
\ba
J^i_{12} = \int \frac{ds_1 d\Gamma_\pm}{\pi} \brm{\Phi^i_{12}}^2,
\qquad
i = e,\pi.
\label{JDefinition}
\ea
The phase volume we parameterize as
(we use the relation
$d^4 q_\pm = \frac{s}{2} d\alpha_\pm d\beta_\pm d^2 q_{\pm\perp}$)
\ba
\frac{1}{\pi} ds_1 d\Gamma_\pm=
\frac{1}{\pi}ds_1\frac{d^3q_+}{2E_+}\frac{d^3q_-}{2E_-}\delta^4(q_1+q_2-q_+-q_-)=\frac{dx_+d^2q_+}{2\pi x_+x_-}.
\label{PairPhaseVolume}
\ea
And performing the phase volume integration we obtain
(see details in Appendix~\ref{AppendixImpactFactor}):
\ba
J^e_{12} &=& 4 \int\limits_0^1 dx \int\limits_0^1 \frac{dt}{D}
\brs{\vec q_1^2\vec q_2^2 \br{a+b-4a b} - 2ab \br{\vec q_1\vec q_2}^2},
\label{JE}
\\
D&=& m^2 + a \vec q_1^2 + b \vec q_2^2, \nn
\ea
where $a=t(1-t)$ and $b=x(1-x)$. After averaging over $\vec q_1$ and $\vec q_2$ vectors
orientations we get the well known result \cite{Frolov:1970ij, Gribov:1970ik, Lipatov:1971ax}:
\ba
\bar J^e_{12} &=& 4 \vec q_1^2\vec q_2^2
\int\limits_0^1 \frac{dx dt}{D}
\brs{a+b-5a b}.
\label{JEAv}
\ea
In case of pion pair production we have:
\ba
J^\pi_{12} &=& \frac{1}{2} \int\limits_0^1 \frac{dx dt}{D}
\brs{\vec q_1^2\vec q_2^2 \br{1-4a}\br{1-4b} + 8ab \br{\vec q_1\vec q_2}^2},
\label{JPi}
\\
\bar J^\pi_{12} &=& \frac{1}{2} \vec q_1^2\vec q_2^2
\int\limits_0^1 \frac{dx dt}{D}
\brs{1-4a-4b+20ab}.
\nn
\ea

We also need to consider the sub-process of pair production by one real and one virtual
photons
$\gamma(k)+\gamma^*(\chi)\to e^+(q_+)+e^-(q_-)$ and
$\gamma(k)+\gamma^*(\chi)\to \pi^+(q_+)+\pi^-(q_-)$.
The values similar to (\ref{InitialPhiE}) and (\ref{InitialPhiPi}) in that case reads as:
\ba
\Phi_0^e &=&
\frac{1}{s}\bar{u}(q_-)
\br{\hat{e}\frac{-\hat{q}_++\hat{\chi}+m}{(q_--k)^2-m^2}\hat{p}+
\hat{p}\frac{\hat{q}_--\hat{\chi}+m}{(-q_+-k)^2-m^2}\hat{e}}
v(p_+),\nn
\label{InitialPhiE0} \\
\Phi_0^\pi &=&
\frac{1}{s}\br{\frac{(2q_--k)e- (-2q_++\chi)p}{(q_--k)^2-m^2}+
\frac{(-2q_++k)e- (2q_--\chi)p}{(-q_++k)^2-m^2}-2ep},
\label{InitialPhiPi0}
\ea
where $e$ is the initial photon polarization 4-vector
$ee^*=-\vec{e}\vec{e}^*=-1$, $ep=ek=0$.
Again, using Dirac equation we can rewrite them in form:
\ba
\Phi_0^e &=&
\bar{u}(q_-)\brs{x_+x_-\br{\frac{1}{D_-}-\frac{1}{D_+}}\hat e
-\frac{x_-}{sD_-}\hat{e}\hat{\chi}\hat{p}+
\frac{x_+}{sD_+}\hat{p}\hat{\chi}\hat{e}}v(q_+),
\label{Phi0E}\\
\Phi_0^\pi &=& -2x_+x_-\brs{\frac{\vec{q}_-\vec{e}}{D_-}+\frac{\vec{q}_+\vec{e}}{D_+}},
\qquad
D_\pm=\vec{q}_\pm^2+m^2.
\label{Phi0Pi}
\ea
So now we define the values similar to (\ref{JDefinition})
in case of one real photon:
\ba
I^{eA}_{01} &=& \int \frac{ds_1d\Gamma_\pm}{\pi} \brm{\Phi_0^e}^2 =
\frac{2}{3}\int\limits_0^1 \frac{dt}{D_0}
\br{\vec{e}_A\vec{e}_A^*\vec{\chi}^2 (1+2a)-2a(\vec{e}_A\vec{\chi})(\vec{e}_A^*\vec{\chi})},
\label{IEImpactFactor}\\
I^{\pi A}_{01} &=& \int \frac{ds_1 d\Gamma_\pm}{\pi} \brm{\Phi_0^\pi}^2 =
\frac{1}{6}\int\limits_0^1 \frac{dt}{D_0}
\br{\vec{e}_A\vec{e}_A^*\vec{\chi}^2(1-4a)+4a(\vec{e}_A\vec{\chi})(\vec{e}_A^* \vec{\chi})},
\label{IPiImpactFactor}
\ea
where $D_0=m^2+a\vec{\chi}^2$ and $a=t(1-t)$. The similar expression can be written down
for the case when particle $B$ is a photon with polarization vector $e_B$: $e_A \to e_B$.

Having this boundary impact factors (\ref{IEImpactFactor}) and (\ref{IPiImpactFactor})
we can write the cross sections of production of two pairs of charged
particles in two polarized photons collision:
\ba
\sigma^{\gamma\gamma}_{ij}=\frac{\alpha^4}{\pi}\frac{d^2\vec{q}}{(\vec{q}^2)^2}
I^j(\vec{q},\vec{e}_2)I^i(\vec{q},\vec{e}_1),
\qquad i,j=e,\pi.
\label{Impact0}
\ea
This result is in agreement with Cheng-Wu impact picture of high-energy
collisions \cite{Cheng:1969bf, Cheng:1969eh, Cheng:1970ef}.
Different distributions based on (\ref{Impact0}) was considered in \cite{Kuraev:1983tf, Kuraev:1985}.

To obtain this formula we had rearranged the phase volume of final particles introducing new
variables -- the transferred momentum $q=\alpha p+\beta k+q_\bot$
and invariant masses of pairs $s_1=s\alpha$, $s_2=-s\beta$:
\ba
d\Gamma^{\gamma\gamma}_{2e^+2e^-} &=&
\br{2\pi}^{-8}
\frac{d^3 q_{1+}d^3 q_{1-}}{2 E_{1+} 2 E_{1-}}
\frac{d^3 q_{2+}d^3 q_{2-}}{2 E_{2+} 2 E_{2-}}
\delta^4\br{k + p - q_{1+} - q_{1-} - q_{2+} - q_{2-}}
\times \nn\\
&\times&
d^4 q
\delta^4\br{p-q-q_{2+}-q_{2-}}
= \nn\\
&=&
\frac{\br{2\pi}^{-8}}{2s}
d\Gamma_{1\pm} ds_1 d\Gamma_{2\pm} ds_2 d^2 \vec q,
\qquad
s_1 = s\alpha,
\qquad
s_2 = -s\beta,
\ea
where we used the known expression $d^4q=(1/(2s))d^2 \vec q ds_1 ds_2$.

In case of unpolarized photons the cross section of production of two
electron-positron pairs is
\ba
\sigma_{2e^+2e^-}^{\gamma\gamma}=\frac{\alpha^4}{\pi m^2}J_e,
\ea
where
\ba
J_e&=&\frac{4}{9}\int\limits_0^\infty dz
\br{\int\limits_0^1 dx \frac{1+x(1-x)}{1+z x(1-x)}}^2 =
\frac{1}{36}\brs{175\zeta_3-38} \approx 4.78,
\label{JTwoPairsDefinition}
\ea
where $\zeta_3=\sum_{k=1}^\infty \frac{1}{k^3}\approx 1.202...$.
In case of production of two $\pi^+\pi^-$ pairs we have:
\ba
\sigma_{2\pi^+2\pi^-}^{\gamma\gamma}=\frac{\alpha^4}{\pi M_\pi^2} J_\pi,
\ea
where
\ba
J_\pi&=&\frac{1}{36}\int\limits_0^\infty dz
\br{\int\limits_0^1 dx \frac{1-2 x(1-x)}{1+z x(1-x)}}^2 =
\frac{1}{144}\brs{7 \zeta_3 + 10} \approx 0.128.
\label{JTwoPoinPairsDefinition}
\ea

\section{Neutral pions production}

Consider now the neutral pion production
in $\gamma\gamma$-collisions. We will obtain below the
cross sections of one, two and a few neutral pions production.

For completeness we put here the results for subprocess
$\gamma^* \gamma^* \to \pi^0$ and
relevant cross sections.

First let us consider the base decay
of neutral pion to two real photons $\pi^0 \to 2\gamma$. It's amplitude
have a form:
\ba
M\br{\pi^0 \to \gamma\br{k_1,e_1}\gamma\br{k_2,e_2}}
=\frac{\alpha}{\pi f_\pi} \br{k_1 e_1 k_2 e_2},
\qquad
f_\pi = 94\MeV,
\ea
where we used notation
$\br{abcd} = \varepsilon_{\alpha\beta\gamma\delta} a^\alpha b^\beta c^\gamma d^\delta$.
This amplitude gives the well known result:
\ba
\Gamma_{\pi^0 \to 2\gamma} =
\frac{\alpha^2 M_\pi^3}{64 \pi^3 f_\pi^2} = 7.2\eV.
\ea
Thus for process $\gamma^* \gamma^* \to \pi^0$ we obtain:
\ba
M_{12}\br{\gamma^*\br{q_1,e_1}\gamma^*\br{q_2,e_2}\to \pi^0}
=\frac{\alpha}{\pi f_\pi} \br{q_1 e_1 q_2 e_2} R_{12},
\ea
where we took into account the external momentum dependence as:
\ba
R_{12} = R_{21} =
2
\int\limits_0^1 d x_1
\int\limits_0^{1-x_1} d x_2
\frac{M^2}{M^2 - M_\pi^2 x_2 x_3 + \vec q_1^2 x_1 x_2 + \vec q_2^2 x_1 x_3},
\ea
where $x_3 = 1-x_1-x_2$ and $M$ is the constituent quark mass in the
fermion triangle loop. The quantity $R_{12}$ is normalized as
$\left. R_{12} \right|_{M^2 \gg M_\pi^2, \vec q_i^2} = 1$. If one of photons is real, then:
\ba
&&M_{01}\br{\gamma^*\br{q_1,e_1}\gamma\br{q_2,e_2}\to \pi^0}
=\frac{\alpha}{\pi f_\pi} \br{q_1 e_1 q_2 e_2} R_{01}, \nn \\
&&R_{01} = R_{10} =
2
\int\limits_0^1 d x_1
\int\limits_0^{1-x_1} d x_2
\frac{M^2}{M^2 - M_\pi^2 x_2 x_3 + \vec q_1^2 x_1 x_2}. \nn
\ea

So now we can consider the processes 
\ba
\gamma(k,e_1)+\gamma(p,e_2) &\to& \pi^0(p_1) + \pi^0(p_2), \nn\\
\gamma(k,e_1)+\gamma(p,e_2) &\to& \pi^0(p_1) + \pi^0(p_2) + \pi^0(p_3), \nn\\
\gamma(k,e_1)+\gamma(p,e_2) &\to& \pi^0(p_1) + \pi^0(p_2) + \cdots + \pi^0(p_{n+2}). \nn
\ea
The phase volumes of final states are:
\ba
d\Gamma^{2\pi^0} &=&
\frac{d^3 p_1}{2E_1}\frac{d^3 p_2}{2E_2}
\br{2\pi}^{4-3 \cdot 2}
\delta^4\br{p+k-p_1-p_2} = \nn\\
&=& \br{2\pi}^{-2} \frac{d^2 q}{2s}, \\
d\Gamma^{3\pi^0} &=&
\frac{d^3 p_1}{2E_1}\frac{d^3 p_2}{2E_2}\frac{d^3 p_3}{2E_3}
\br{2\pi}^{4-3 \cdot 3}
\delta^4\br{p+k-p_1-p_2-p_3} = \nn\\
&=& \br{2\pi}^{-5} \frac{d\beta_1}{\beta_1}
\frac{d^2 q_1 d^2 q_2}{4s}, \qquad
\frac{M_\pi^2}{s} \ll \beta_1 \ll 1,\\
d\Gamma^{(n+2)\pi^0} &=&
\frac{d^3 p_1}{2E_1}\frac{d^3 p_2}{2E_2} \cdots \frac{d^3 p_{n+2}}{2E_{n+2}}
\br{2\pi}^{4-3 \cdot (n+2)}
\delta^4\br{p+k-p_1-p_2-\cdots-p_{n+2}} = \nn\\
&=& \br{2\pi}^{4-3(n+2)}
\br{\prod_{i=1}^n \frac{d\beta_i}{\beta_i}}
\prod_{i=1}^{n+1} d^2 q_i
\frac{1}{2^{n+1} s}, \nn\\
&&\frac{M_\pi^2}{s} \ll \beta_n \ll \beta_{n-1} \ll \cdots \ll \beta_1 \ll 1.
\label{BetaArrangement}
\ea
The amplitudes of these processes have a form:
\ba
M^{2\pi^0} &=&
\br{\frac{\alpha}{\pi f_\pi}}^2
\frac{s}{2}
\brs{\vec e_1 \vec n} \brs{\vec e_2 \vec n} R_{0q}^2, \\
M^{3\pi^0} &=&
\br{\frac{\alpha}{\pi f_\pi}}^3
\frac{s}{2}
\brs{\vec e_1 \vec n_1} \brs{\vec n_1 \vec n_2} \brs{\vec n_2 \vec e_2}
R_{01} R_{12} R_{20}, \\
M^{(n+2)\pi^0} &=&
\br{\frac{\alpha}{\pi f_\pi}}^{n+2}
\frac{s}{2}
\brs{\vec e_1 \vec n_1} \brs{\vec n_1 \vec n_2} \cdots \brs{\vec n_n \vec n_{n+1}} \brs{\vec n_{n+1} \vec e_2} \times \nn\\
&\times& R_{01} R_{12} \cdots R_{n,n+1} R_{n+1,0},
\ea
\ba
\vec n = \frac{\vec q}{\brm{\vec q}}, \qquad
\vec n_i = \frac{\vec q_i}{\brm{\vec q_i}}, \nn
\ea
\ba
\brs{\vec e \vec n}^2 = 1 - \br{\vec e \vec n}\br{\vec e^* \vec n},
\qquad
\brs{\vec n_1 \vec n_2}^2 = \sin^2 \phi_{12},
\ea
where $\phi_{12}$ is the azimuthal angle between 2-vector
$\vec q_1$ and $\vec q_2$.
The cross sections then reads as:
\ba
d\sigma^{\gamma\gamma\to 2\pi^0}
&=&
\frac{\alpha^4}{2^8 \pi^5 f_\pi^4}
\brs{\vec e_1 \vec n}^2 \brs{\vec e_2 \vec n}^2 R_{0q}^4
\frac{d^2 q}{\pi}, \\
d\sigma^{\gamma\gamma\to 3\pi^0} &=&
\frac{\alpha^6}{2^{12} \pi^9 f_\pi^6}
\brs{\vec e_1 \vec n_1}^2 \brs{\vec n_1 \vec n_2}^2 \brs{\vec n_2 \vec e_2}^2
R_{01}^2 R_{12}^2 R_{20}^2 L_\pi
\frac{d^2 q_1 d^2 q_2}{\pi^2}, \\
d\sigma^{\gamma\gamma\to (n+2)\pi^0} &=&
2^{-4n-8}\pi^{-4n-5} \br{\frac{\alpha^2}{f_\pi^2}}^{n+2}
\brs{\vec e_1 \vec n_1}^2 \brs{\vec n_1 \vec n_2}^2 \cdots \brs{\vec n_n \vec n_{n+1}}^2 \brs{\vec n_{n+1} \vec e_2}^2 \times \nn\\
&\times& R_{01}^2 R_{12}^2 \cdots R_{n,n+1}^2 R_{n+1,0}^2
\frac{L_\pi^n}{n!}
\br{\prod_{i=1}^{n+1}\frac{d^2 q_i}{\pi}},
\ea
here $L_\pi=\ln \frac{s}{M_\pi^2}$.
We also can consider the processes $\gamma e \to (n \pi^0) e$ and
$e e \to e e (n \pi^0)$ in a same manner.

In case of photo-production of $(n+2)$ pions on electron:
\ba
\gamma(k,e)+e(p) &\to& \pi^0(p_1) + \pi^0(p_2) + \cdots + \pi^0(p_{n+2}) + e(p'), \nn
\ea
matrix element have a form:
\ba
M^{\gamma e \to e (n+2)\pi^0} &=&
\br{4\pi\alpha}^{1/2}
\br{\frac{\alpha}{\pi f_\pi}}^{n+2}
s N_e
\brs{\vec e \vec n_1} \brs{\vec n_1 \vec n_2} \cdots \brs{\vec n_{n+1} \vec n_{n+2}} \times \nn\\
&\times& R_{01} R_{12} \cdots R_{n+1,n+2}
\frac{\brm{\vec q_{n+2}}}{q_{n+2}^2},
\ea
the phase volumes of final state is:
\ba
d\Gamma^{\gamma e \to e(n+2)\pi^0} &=&
\frac{d^3 p_1}{2E_1}\frac{d^3 p_2}{2E_2} \cdots \frac{d^3 p_{n+2}}{2E_{n+2}}
\frac{d^3 p'}{2E'}
\br{2\pi}^{4-3 \cdot (n+3)} \times \nn\\
&\times&\delta^4\br{p+k-p'-p_1-p_2-\cdots-p_{n+2}} = \nn\\
&=& \br{2\pi}^{4-3(n+3)}
\br{\prod_{i=1}^{n+1} \frac{d\beta_i}{\beta_i}}
\prod_{i=1}^{n+2} \frac{d^2 q_i}{\pi}
\frac{1}{2^{n+2} s} \pi^{n+2},
\ea
and the cross section then reads as:
\ba
d\sigma^{\gamma e\to e(n+2)\pi^0} &=&
\alpha
\br{\frac{\alpha^2}{f_\pi^2}}^{n+2}
\frac{\pi}{\br{2 \pi}^{4n+7}}
\br{\prod_{i=1}^{n+1}\frac{d^2 q_i}{\pi}}
\frac{\vec q_{n+2}^2 \br{d^2 q_{n+2}/\pi}}
{\brs{\vec q_{n+2}^2 + m_e^2 \br{\frac{M_\pi^4}{s^2\beta_{n+1}^2}}}^2}
\times\nn\\
&\times&
\brs{\vec e_1 \vec n_1}^2 \brs{\vec n_1 \vec n_2}^2 \cdots \brs{\vec n_{n+1} \vec n_{n+2}}^2
R_{01}^2 R_{12}^2 \cdots R_{n+1,n+2}^2
\br{\prod_{i=1}^{n+1}\frac{d\beta_i}{\beta_i}}.
\ea
In particular in case of one and two pions production we have:
\ba
d\sigma^{\gamma e\to e \pi^0} &=&
\frac{\alpha^3}{f_\pi^2}
\frac{1}{8\pi^2}
\frac{d^2 q}{\pi}
\frac{\vec q_n^2}
{\brs{\vec q^2 + m_e^2 \br{\frac{M_\pi^2}{s}}^2}^2}
\brs{\vec e \vec n}^2 R_{01}^2, \\
d\sigma^{\gamma e\to e 2\pi^0} &=&
\frac{\alpha^5}{f_\pi^4}
\frac{1}{2^7 \pi^6}
\frac{d^2 q_1 d^2 q_2}{\pi^2}
\frac{\vec q_2^2}
{\brs{\vec q_2^2 + m_e^2 \br{\frac{M_\pi^2}{s\beta_1}}^2}^2}
\brs{\vec e \vec n_1}^2 \brs{\vec n_1 \vec n_2}^2
R_{01}^2 R_{12}^2 \frac{d\beta_1}{\beta_1}.
\ea
In particular the single pion production total cross section is
(Primakoff process):
\ba
\sigma^{\gamma e\to e \pi^0}(s) &=&
\frac{\alpha^3}{4\pi^2 f_\pi^2}
\br{1-\br{\vec e \vec n}\br{\vec e^* \vec n}}
\ln \frac{s}{m_e M_\pi}.
\ea

In case of production of $n$ pions on electron-electron collisions:
\ba
e(k)+e(p) &\to& e(k') + e(p') + \pi^0(p_1) + \pi^0(p_2) + \cdots + \pi^0(p_n), \nn
\ea
the matrix element have a form:
\ba
M^{e e \to e e (n\cdot \pi^0)} &=&
2 s N_1 N_2 \br{4\pi\alpha}
\br{\frac{\alpha}{\pi f_\pi}}^n
\br{\prod_{i=1}^n \brs{\vec n_i \vec n_{i+1}} R_{i,i+1}}
\ea
the phase volumes of final state is:
\ba
d\Gamma^{e e \to e e (n \cdot \pi^0)} &=&
\frac{d^3 k'}{2E_k'} \frac{d^3 p'}{2E_p'}
\frac{d^3 p_1}{2E_1}\frac{d^3 p_2}{2E_2} \cdots \frac{d^3 p_n}{2E_n}
\br{2\pi}^{4-3 \cdot (n+2)} \times \nn\\
&\times&\delta^4\br{p+k-p'-k'-p_1-p_2-\cdots-p_n} = \nn\\
&=& \br{2\pi}^{4-3(n+2)}
\frac{1}{2^{n+1} s} \pi^{n+1}
\prod_{i=1}^{n+1} \frac{d^2 q_i}{\pi}
\prod_{i=1}^n \frac{d\beta_i}{\beta_i},
\ea
and the cross section then reads as:
\ba
d\sigma^{e e \to e e (n \cdot \pi^0)} &=&
\frac{\alpha^{2(1+n)}}{2^{4n-2} \pi^{4n-1} f_\pi^{2n}}
\br{\prod_{i=2}^n\frac{d^2 q_i}{\pi}}
\br{\prod_{i=1}^n \brs{\vec n_i \vec n_{i+1}}^2 R_{i,i+1}^2}
\times \nn\\
&\times&
\frac{d^2 q_1}{\pi}
\frac{\vec q_1^2}
{\brs{\vec q_1^2 + m_1^2 \beta_1^2}^2}
\frac{d^2 q_{n+1}}{\pi}
\frac{\vec q_{n+1}^2}
{\brs{\vec q_{n+1}^2 + m_2^2 \alpha_{n+1}^2}^2}
\br{\prod_{i=1}^n\frac{d\beta_i}{\beta_i}},
\ea
where $\alpha_{n+1} = M_\pi^2/(s \beta_n)$.
In particular in case of one and two pions production we have:
\ba
d\sigma^{e e\to e e \pi^0} &=&
\frac{\alpha^4}{4\pi^3 f_\pi^2}
\frac{d^2 q_1}{\pi}
\frac{\vec q_1^2}
{\brs{\vec q_1^2 + m_1^2 \beta_1^2}^2}
\frac{d^2 q_2}{\pi}
\frac{\vec q_2^2}
{\brs{\vec q_2^2 + m_2^2 \alpha_2^2}^2}
\brs{\vec n_1 \vec n_2}^2 R_{12}^2
\frac{d\beta_1}{\beta_1}, \\
d\sigma^{e e\to e e 2\pi^0} &=&
\frac{\alpha^6}{2^6 \pi^7 f_\pi^4}
\frac{d^2 q_1}{\pi}
\frac{\vec q_1^2}
{\brs{\vec q_1^2 + m_1^2 \beta_1^2}^2}
\frac{d^2 q_3}{\pi}
\frac{\vec q_3^2}
{\brs{\vec q_3^2 + m_2^2 \alpha_3^2}^2}
\times \nn\\
&\times&
\brs{\vec n_1 \vec n_2}^2 \brs{\vec n_2 \vec n_3}^2
\frac{d^2 q_2}{\pi}
R_{12}^2 R_{23}^2
\frac{d\beta_1}{\beta_1}
\frac{d\beta_2}{\beta_2}.
\ea

In general for process $e+e \to e+e+T$ with $T$ is neutral pion or a single pair in case then the
transfers momenta to the projectiles do not exceed
some values $Q_1,Q_2$
$\sqrt{-q_1^2}<Q_1$ and $\sqrt{-q_2^2}<Q_2$ the integral in r.h.s. can be performed with logarithmic accuracy:
\ba
\int\limits_{M_T^2/s}^1\frac{d\beta_1}{\beta_1}
\ln \br{\frac{Q_1^2}{m_e^2 \beta_1^2}}
\ln \br{\frac{Q_2^2}{m_e^2 \alpha_2^2}} =
\frac{2}{3}[L_T^2 + 3 L_T \ln\frac{Q_1 Q_2}{m_e^2} +
6 \ln\frac{Q_1}{m_e}\ln\frac{Q_2}{m_e}],
\ea
where $L_T = \ln\br{s/M_T^2}$, $s\alpha_2\beta_1 = M_T^2$ and $M_T$ is the characteristic mass of
produced system $T$. In case of processes for process $p+p \to p+p+T$ the
$m_e \to M_p$ replacement should be done.
In particular the total cross section of single neutral pion production is
\ba
\sigma^{e e\to e e \pi^0} &=&
\frac{\alpha^4 L_\pi}{12 \pi^3 f_\pi^2}
\brs{L_\pi^2 + 6 l \br{L_\pi+l}}, \qquad Q_1=Q_2<M_\pi,
\label{Tot}
\ea
where $l = \ln \br{M_\pi/m_e}$.

\section{The multiple pair production}
Let us now consider the processes of multiple pairs production at electron-positron,
electron-photon and photon-photon colliders.

\subsection{Process $\gamma\gamma \to n(e^+e^-)\mbox{ or }n(\pi^+\pi^-)$}

The process of $n$ pairs production at photon-photon collisions are shown on
Fig.~\ref{FigGammaGammaNPairs} and the matrix elements of two, three and $n$ pairs
production is shown below:
\begin{figure}
\includegraphics{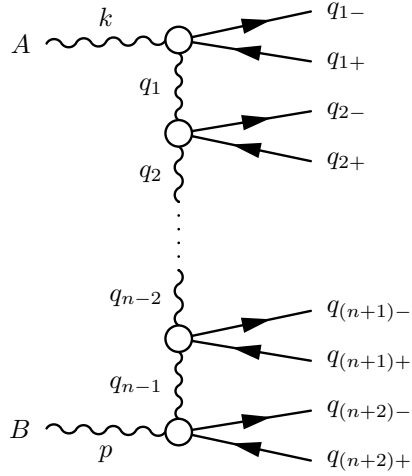}
\caption{Feynman diagram for $n$ pions pairs production at photon-photon collisions.}
\label{FigGammaGammaNPairs}
\end{figure}
\ba
M^{(\gamma\gamma)}_{(2)} &=&
\frac{\br{4\pi\alpha}^2 s}{q_1^2} 2 \Phi_0^A \Phi_0^B, \nn\\
M^{(\gamma\gamma)}_{(3)} &=&
\frac{\br{4\pi\alpha}^3 s}{q_1^2 q_2^2} 2^2 \Phi_0^A \Phi_{12} \Phi_0^B, \nn\\
\cdots \nn\\
M^{(\gamma\gamma)}_{(2+n)} &=&
\frac{\br{4\pi\alpha}^{n+2} s}{q_1^2 \cdots q_{n+1}^2} 2^{n+1} \Phi_0^A \Phi_{12} \cdots \Phi_{n,n+1} \Phi_0^B,
\ea
where $\Phi_0^A$ and $\Phi_0^B$ are boundary factors
(see (\ref{Phi0E}), (\ref{Phi0Pi})),
which describe pair top and bottom pairs production
from one real and one virtual photons.
The indexes $A$ and $B$ defines the type of particles which are produced.
Factors $\Phi_{ij}$
(see (\ref{PhiE}), (\ref{InitialPhiPi})) describe pair production in collision of two
virtual photons with momenta $q_i$ and $q_j$.
Factor $2^{n+1}$ comes from Gribov's representation of photon
propagator (\ref{GribovRepresentation}).

The phase volumes for two, three, and $(n+2)$ pairs production
have a form (by analogy with the case of two pairs
production, considered above, we introduce
one, two, and $(n+1)$ transferred momenta):
\ba
d\Gamma^{(\gamma\gamma)}_{(2)} &=&
2^{-9}\pi^{-5} \frac{1}{s}
\br{\frac{ds_1 d\Gamma^\pm_1}{\pi}\frac{ds_2 d\Gamma^\pm_2}{\pi}}
\frac{d\vec q_1}{\pi} , \nn\\
d\Gamma^{(\gamma\gamma)}_{(3)} &=&
2^{-16}\pi^{-9} \frac{1}{s}
\br{\frac{ds_1 d\Gamma^\pm_1}{\pi}\frac{ds_2 d\Gamma^\pm_2}{\pi}\frac{ds_3 d\Gamma^\pm_3}{\pi}}
\frac{d\vec q_1}{\pi} \frac{d\vec q_2}{\pi}\frac{d\beta_1}{\beta_1}, \nn\\
\cdots \nn\\
d\Gamma^{(\gamma\gamma)}_{(2+n)} &=&
2^{-7n-9} \pi^{-4n-5} \frac{1}{s}
\br{\prod_{i=1}^{n+2} \frac{ds_i d\Gamma^\pm_i}{\pi}}
\br{\prod_{i=1}^{n+1} \frac{d\vec q_i}{\pi}} \br{\prod_{i=1}^{n} \frac{d\beta_i}{\beta_i}},
\ea
where $s_i$ and $d\Gamma^\pm_i$ are the invariant mass and phase volume of $i$-th pair
(\ref{PairPhaseVolume}).

And the cross section for $(2+n)$ pairs production have a form:
\ba
\sigma^{(\gamma\gamma)}_{(2+n)} &=&
\frac{1}{8s} \int \brm{M^{(\gamma\gamma)}_{(2+n)}}^2 d\Gamma^{(\gamma\gamma)}_{(2+n)} = \nn\\
&=&
\frac{\alpha^{2(n+2)}}{2^{n+2} \pi^{2n+1}} \int
\br{\prod_{i=1}^{n+1} \frac{d\vec q_i}{\pi \br{\vec q_i^2}^2}} \br{\prod_{i=1}^{n} \frac{d\beta_i}{\beta_i}}
I^A_{01} J_{12} J_{23} \cdots J_{n,n+1} I^B_{n+1,0},
\label{CrossSectionGammaGammatoNpairs}
\ea
where we entered the following impact factors:
\ba
I^A_{01} &=& \int \frac{s_1 d\Gamma^\pm_1}{\pi} \brm{\Phi^A}^2, \qquad
J_{ij} = \int \frac{s_i d\Gamma^\pm_i}{\pi} \brm{\Phi_{ij}}^2, \nn \\
I^B_{n+1,0} &=& \int \frac{s_{n+2} d\Gamma^\pm_{n+2}}{\pi} \brm{\Phi^B}^2. \nn
\ea

The explicit form of them was already calculated before
(see (\ref{JE}), (\ref{JEAv}), (\ref{JPi}), (\ref{IEImpactFactor}), (\ref{IPiImpactFactor})).

\subsection{Process $\gamma e \to n(e^+e^-) e\mbox{ or }n(\pi^+\pi^-) e$}

The process of $n$ pairs production at electron-photon collisions are shown on
Fig.~\ref{FigGammaENPairs} and the matrix elements of two, three and $2+n$ pairs
production is shown below:
\begin{figure}
\includegraphics{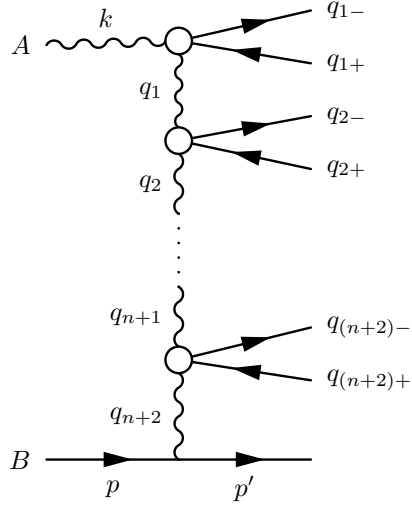}
\caption{Feynman diagram for $n+2$ pairs production at electron-photon collisions.}
\label{FigGammaENPairs}
\end{figure}
\ba
M^{(\gamma e)}_{(1)} &=&
\frac{\br{4\pi\alpha}^{\frac{3}{2}} s}{q_1^2} 2 N_B \Phi_0^{A}, \nn\\
M^{(\gamma e)}_{(2)} &=&
\frac{\br{4\pi\alpha}^{\frac{5}{2}} s}{q_1^2 q_2^2} 2^2 N_B \Phi_0^{A} \Phi_{12}, \nn\\
\cdots \nn\\
M^{(\gamma e)}_{(2+n)} &=&
\frac{\br{4\pi\alpha}^{\frac{2n+5}{2}} s}{q_1^2 \cdots q_{n+2}^2} 2^{n+2} N_B \Phi_0^{A}
\Phi_{12}\cdots \Phi_{n+1,n+2},
\ea
where $N_B$ is the boundary factor, which describe bottom electron line.
The phase volumes for one, two, and $(n+2)$ pairs production
have a form: (again we introduce the set of transferred momenta)
\ba
d\Gamma^{(\gamma e)}_{(1)} &=&
2^{-6}\pi^{-3} \frac{1}{s}
\br{\frac{ds_1 d\Gamma^\pm_1}{\pi}}
\frac{d\vec q_1}{\pi}, \nn\\
d\Gamma^{(\gamma e)}_{(2)} &=&
2^{-13}\pi^{-7} \frac{1}{s}
\br{\frac{ds_1 d\Gamma^\pm_1}{\pi}\frac{ds_2 d\Gamma^\pm_2}{\pi}}
\frac{d\vec q_1}{\pi} \frac{d\vec q_2}{\pi}\frac{d\beta_1}{\beta_1}, \nn\\
\cdots \nn\\
d\Gamma^{(\gamma e)}_{(2+n)} &=&
2^{-7n-13} \pi^{-4n-7} \frac{1}{s}
\br{\prod_{i=1}^{n+2} \frac{ds_i d\Gamma^\pm_i}{\pi}}
\br{\prod_{i=1}^{n+2} \frac{d\vec q_i}{\pi}} \br{\prod_{i=1}^{n+1} \frac{d\beta_i}{\beta_i}},
\ea
and the cross section for $(2+n)$ pairs production have a form:
\ba
\sigma^{(\gamma e)}_{(2+n)} &=&
\frac{1}{8s} \int \sum \brm{M^{(\gamma e)}_{(2+n)}}^2 d\Gamma^{(\gamma e)}_{(2+n)} = \nn\\
&=&
\frac{\alpha^{2n+5)}}{2^n \pi^{2(n+1)}}\int
\br{\prod_{i=1}^{n+1} \frac{d\vec q_i}{\pi \br{\vec q_i^2}^2}}
\br{\prod_{i=1}^{n+1} \frac{d\beta_i}{\beta_i}}
\br{2\ln\br{\frac{s\beta_{n+1}}{m^2}}} \times \nn\\
&\times&
I^A_{01} J_{12} \cdots J_{n,n+1} I_{n+1,0}, \qquad
I_{n+1,0}=\br{\frac{J_{n+1,n+2}}{\vec{q}_{n+2}^2}}_{\vec{q}_{n+2}=0}.
\label{CrossSectionGammaEtoNpairs}
\ea
For processes $\gamma e\to e T$ the Weizsaecker-Williams enhancement mechanism works.
As a result the factors $\ln(1/\alpha_{n+2}^2)\approx \ln(s\beta_{n+1}/M_T^2)^2$,
$\ln(1/\beta^2)$ appears.
For details see Appendix \ref{AppendixWW}.

In particular cases of creation of one and two fermion pairs we obtain
(in agreement with Lipatov and Frolov result \cite{Frolov:1970ij, Gribov:1970ik, Lipatov:1971ax}):
\ba
\sigma^{\gamma e\to e^+e^- e}&=&\frac{28}{9}\frac{\alpha^3}{m^2}L_e, \nn \\
\sigma^{\gamma e\to 2e^+2e^- e}&=&\frac{\alpha^5}{\pi^2 m^2} L_e^2 J_e,
\qquad Q_2<m,
\ea
where $L_e=\ln\br{s/m_e^2}$ and $J_e$ was defined in (\ref{JTwoPairsDefinition}).
In cases of one and two charged pion pairs production we obtain:
\ba
\sigma^{\gamma e\to \pi^+\pi^- e}&=&\frac{2 \alpha^3}{9M_\pi^2} \ln\br{\frac{s}{m M_\pi}},
\qquad Q_2 < M_\pi, \nn \\
\sigma^{\gamma e\to 2\pi^+2\pi^- e}&=&\frac{\alpha^5}{\pi^2 M_\pi^2} L_\pi L_e J_\pi,
\qquad Q_2<M_\pi,
\ea
where $L_\pi = \ln\br{s/M_\pi^2}$ and $J_\pi$ was defined in (\ref{JTwoPoinPairsDefinition}).

\subsection{Process $e^+ e^- \to n(e^+e^-)\mbox{ or }n(\pi^+\pi^-)$}

The process of $n$ pairs production at electron-positron collisions are shown on
Fig.~\ref{FigEENPairs}. The matrix elements of one and $(2+n)$ pairs
production is shown below:
\begin{figure}
\includegraphics{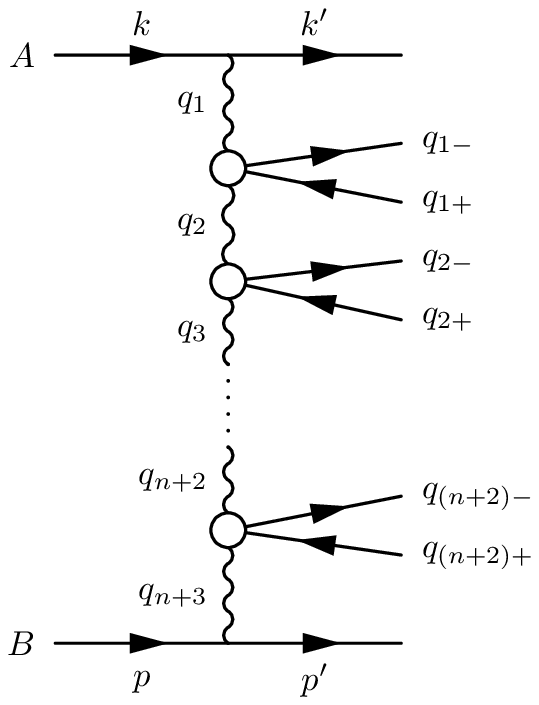}
\caption{Feynman diagram for $n+2$ pairs production at electron-positron collisions.}
\label{FigEENPairs}
\end{figure}
\ba
M^{(ee)}_{(1)} &=&
\frac{\br{4\pi\alpha}^2 s}{q_1^2 q_2^2} 2^2 N_A \Phi_{12} N_B, \nn\\
\cdots \nn\\
M^{(ee)}_{(2+n)} &=&
\frac{\br{4\pi\alpha}^{n+3} s}{q_1^2 \cdots q_{n+3}^2} 2^{n+3} N_A
\Phi_{12}\cdots \Phi_{n+2,n+3} N_B,
\ea
where $N_A,N_B$ are lepton's impact factors.
The phase volumes for one, two and $(n+2)$ pairs production
have a form:
\ba
d\Gamma^{(ee)}_{(1)} &=&
2^{-10}\pi^{-5} \frac{1}{s}
\br{\frac{ds_1 d\Gamma^\pm_1}{\pi}}
\frac{d\vec q_1}{\pi} \frac{d\vec q_2}{\pi} \frac{d\beta_1}{\beta_1}, \nn\\
d\Gamma^{(ee)}_{(2)} &=&
2^{-17}\pi^{-9} \frac{1}{s}
\br{\frac{ds_1 d\Gamma^\pm_1}{\pi}\frac{ds_2 d\Gamma^\pm_2}{\pi}}
\frac{d\vec q_1}{\pi} \frac{d\vec q_2}{\pi} \frac{d\vec q_3}{\pi}
\frac{d\beta_1}{\beta_1}\frac{d\beta_2}{\beta_2}, \nn\\
\cdots \nn\\
d\Gamma^{(ee)}_{(2+n)} &=&
2^{-7n-17}\pi^{-4n-9} \frac{1}{s}
\br{\prod_{i=1}^{n+2} \frac{ds_i d\Gamma^\pm_i}{\pi}}
\br{\prod_{i=1}^{n+3} \frac{d\vec q_i}{\pi}}
\br{\prod_{i=1}^{n+2} \frac{d\beta_i}{\beta_i}},
\ea
and the cross section for $(2+n)$ pairs production have a form
\ba
\sigma^{(ee)}_{(2+n)} &=&
\frac{1}{8s} \int \sum \brm{M^{(ee)}_{(2+n)}}^2 d\Gamma^{(ee)}_{(2+n)} = \nn\\
&=&
\frac{\alpha^{2(n+3)}}{2^{n-2} \pi^{2n+3}}\int
\br{\prod_{i=1}^{n+1} \frac{d\vec q_i}{\pi \br{\vec q_i^2}^2}}
\br{\prod_{i=1}^{n+2} \frac{d\beta_i}{\beta_i}}
\br{2\ln\br{\frac{1}{\beta_1}}}
\br{2\ln\br{\frac{s\beta_{n+2}}{m^2}}}
\times\nn\\
&\times&
I_{01} \br{\prod_{i=1}^{n+1} J_{i,i+1}} I_{n+2,0}.
\ea
For the case of creation one and two electron-positron pairs we reproduce the known results
(see \cite{Landau:1934zj, Cheng:1969bf, Cheng:1969eh, Cheng:1970ef, Frolov:1970ij, Gribov:1970ik, Lipatov:1971ax, Racah:1936}):
\ba
\sigma^{ee\to ee e^+e^-}&=&\frac{28}{27}\frac{\alpha^4L_e^3}{\pi m^2}, \nn \\
\sigma^{ee\to ee2e^+2e^-}&=&\frac{\alpha^6L_e^4}{6\pi^3 m^2}J_e,
\qquad Q_1 \sim Q_2 < m.
\label{CrossSectionEEtoNpairs}
\ea
For processes $e\bar e \to e\bar{e} \pi^+\pi^-$ and
$e\bar e \to e\bar{e} 2\pi^+ 2 \pi^-$ we obtain:
\ba
\sigma^{e\bar e \to e\bar{e} \pi^+\pi^-}&=&
\frac{4}{27}\frac{\alpha^4 L_\pi}{M_\pi^2}
[L_\pi^2 + 6l(L_\pi+l)],
\qquad Q_1 \sim Q_2 < m,
 \nn \\
\sigma^{e\bar e \to e\bar{e} 2\pi^+ 2 \pi^-}&=&
\frac{\alpha^6 L_\pi^2}{6 \pi^3 M_\pi^2}
\br{L_\pi^2 + 8 l L_\pi + 12 l^2 }
J_\pi,
\qquad Q_1 \sim Q_2 < m, \nn
\ea
where $L_\pi=\ln\br{s/M_\pi^2}$ and $l=\ln\br{M_\pi/m_e}$.

\section{Generalization}

Let us pay some attention to obtained cross sections
(\ref{CrossSectionGammaGammatoNpairs}),
(\ref{CrossSectionGammaEtoNpairs}),
(\ref{CrossSectionEEtoNpairs}).
They have similar form. The integration over $\beta_i$ leads to some
logarithmical enhancement. Let us demonstrate this in case of electron-positron pairs
production. Keeping in mind the relative arrangements of $\beta_i$
(see (\ref{BetaArrangement})) we obtain:
\ba
\int \prod_{i=1}^{n} \frac{d\beta_i}{\beta_i}
=
L_e^n
\int\limits_0^1 dx_1
\int\limits_0^{x_1} dx_2
\cdots
\int\limits_0^{x_{n-1}} dx_n
= \frac{L_e^n}{n!}.
\ea
This can be presented as the integral over complex variable $j$:
\ba
\frac{L^n}{n!} =
\oint \limits_C \frac{dj}{2\pi i} \frac{\br{s/m^2}^j}{j^{n+1}},
\label{OintRepresentation}
\ea
where contour $C$ is the small circle around the point $j=0$ in the $j$ complex plane.
%
%
Denoting the common part as:
\ba
Z_n =
\frac{\alpha^{2n}}{2^n \pi^{2n} j^n}
\br{\prod_{i=1}^{n} J_{i,i+1}}
\br{\int \prod_{i=1}^{n+1} \frac{d\vec q_i}{\pi \br{\vec q_i^2}^2}},
\ea
we can rewrite these cross sections in form:
\ba
\sigma^{(\gamma\gamma)}_{(2+n)} &=&
\oint \frac{dj}{2\pi i} \br{s/m^2}^j
\br{ \frac{\alpha^4}{4\pi} I^A_{01} Z_n I^B_{n+1,0} }, \nn\\
\sigma^{(\gamma e)}_{(2+n)} &=&
\oint \frac{dj}{2\pi i} \br{s/m^2}^j
\br{ \frac{\alpha^5}{\pi^2 j^2} I^A_{01} Z_n I_{n+1,0} }, \\
\sigma^{(ee)}_{(2+n)} &=&
\oint \frac{dj}{2\pi i} \br{s/m^2}^j
\br{ \frac{4\alpha^6}{\pi^3 j^4} I_{01} Z_n I_{n+1,0} }. \nn
\ea
Considering the production of 1, 2, \dots pairs it is convenient to
introduce the value $\Psi$:
\ba
\Psi = \sum_{n=1}^\infty Z_n =
\sum_{n=1}^\infty
\frac{\alpha^{2n}}{2^n \pi^{2n} j^n}
\br{\prod_{i=1}^{n} J_{i,i+1}}
\br{\int \prod_{i=1}^{n+1} \frac{d\vec q_i}{\pi \br{\vec q_i^2}^2}}.
\ea
Then the cross sections of production of 1, 2, \dots pairs
of charged particles reads as:
\ba
\sigma^{(\gamma\gamma)}_{n\geq 3} &=&
\oint \frac{dj}{2\pi i} \br{s/m^2}^j
\br{ \frac{\alpha^4}{4\pi} I^A_{01} \Psi I^B_{n+1,0} }, \nn\\
\sigma^{(\gamma e)}_{n\geq 1} &=&
\oint \frac{dj}{2\pi i} \br{s/m^2}^j
\br{ \frac{\alpha^5}{\pi^2 j^2} I^A_{01} \Psi I_{n+1,0} }, \label{CSwithArbitraryPairsNumber}\\
\sigma^{(ee)}_{n\geq 1} &=&
\oint \frac{dj}{2\pi i} \br{s/m^2}^j
\br{ \frac{4\alpha^6}{\pi^3 j^4} I_{01} \Psi I_{n+1,0} }. \nn
\ea

\section{kinematics of jets emission}

Let discuss the kinematics of the definite pair production at collision of two virtual photons:
\ba
\gamma^*(q_i)+\gamma^*(-q_{i+1}) \to e^+(q_{+i})+e^-(q_{-i}), i=1,2,...,n. \nn
\ea
Keeping in mind the Sudakov parametrization
of the 4-momentum of pairs component
$q_\pm^i=\alpha_{\pm i} E n_p+\beta_{\pm i} E n_k+q_{\pm\bot i}$ with $2E=\sqrt{s}$
is the center of mass beams total energy.
Light-like 4 vectors $n_p=(1,1,0,0)$, $n_k=(1,-1,0,0)$ have
space component-unit vectors directed along initial 3-momenta
of colliding beams, $\vec{p}$, $\vec{k}$.
We choice a vector $\vec{k}$ as $z$-axes direction.
The transverse to the beam direction component
$q_\bot=(0,0,q_x,q_y)$, $q_\bot^2=-\vec{q}^2$
can be expressed in terms of the polar angle of the relevant
particle emission in the center of mass frame:
\ba
\tan\theta_i=\frac{|\vec{q}_i| }{E(\beta_i-\alpha_i)}, \qquad
\alpha_i=\frac{\vec{q}_i^2+m^2}{4E^2\beta_i}, \qquad
i=\pm.
\ea
Particle with momentum $q_i$ moves in hemisphere $\vec{p}$
if $\alpha_i > \beta_i$ and to hemisphere
$\vec{k}$ in case $\beta_i > \alpha_i$.
We remind here the conservation low $q_i-q_{i+1}=q_{-i}+q_{+i}$.
In Sudakov parametrization it can be written as
\ba
\vec{q}_{\bot,i}-\vec{q}_{\bot,i+1}=\vec{q}_{+i}+\vec{q}_{-i},
\quad
\beta_i=\beta_{+i}+\beta_{-i},
\quad
-\alpha_{i+1}=\alpha_{+i}+\alpha_{-i},
\ea
where we keep into account the rigorous arrangements of longitudinal Sudakov parameters:
\ba
&& 1 \gg \beta_1 \gg \beta_2 \gg \cdots \gg \beta_n \gg \frac{m_1^2}{E^2}, \nn\\
&& 1 \gg \alpha_n \gg \alpha_{n-1} \gg \cdots \gg \alpha_1 \gg \frac{m_2^2}{E^2},
\ea
with $m_{1,2}$ are the masses of colliding particles or the mass of the created pair component in the case when
the relevant initial particle is a photon.

In paper \cite{Kotikov:2000pm} this results were generalized to the
supersymmetric QED case.

\section {Inclusive cross section}
\begin{figure}
\includegraphics[width=0.9\textwidth]{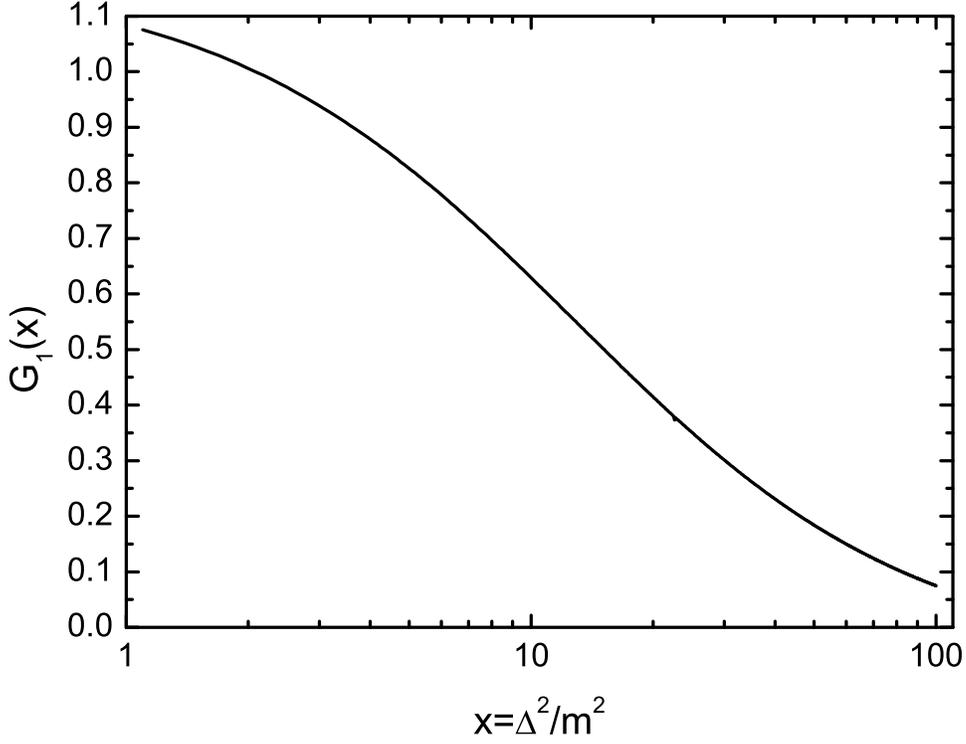}
\caption{Function $G_{1}(x)$ (\ref{G1Definition}) dependence of $x$.}
\label{FigOfG1}
\end{figure}

In the experimental set-up with detecting only one pair emitted in peripheral kinematics
the differential distributions averaged on kinematical parameters of other pairs.
Let us describe the modification of cross section of production of $n$ pairs to obtain the
contribution to inclusive cross section.
Keeping in mind the rigorous arrangement by magnitude of longitudinal Sudakov parameters
we can conclude that two replacements must be done. One of them concern the distribution on the
longitudinal Sudakov parameters of the momentum transferred. Really the replacement must be done:
($m$ is the mass of pair components)
\ba
&&\frac{\delta}{\delta\beta_k}\int\limits_{m^2/s}^1 \frac{d\beta_1}{\beta_1}\int\limits_{m^2/s}^{\beta_1}
\frac{d\beta_2}{\beta_2}...\int\limits_{m^2/s}^{\beta_{n-1}} \frac{d\beta_n}{\beta_n}= \nn \\
&&\qquad=\frac{1}{\beta_k}\ln^{k-1}(1/\beta_k)(L_e-\ln(1/\beta_k))^{n-k}\frac{1}{\Gamma(k)\Gamma(n-k+1)}
\equiv
F_{k,n}(\beta_k).
\ea
The quantity $F_{k,n}(\beta)$ is normalized as
\ba
\int\limits_{m^2/s}^1 d\beta F_{k,n}(\beta)=\frac{L_e^n}{n!}.
\ea
Another one is
\ba
G_{k,n}(\vec{\Delta}^2) \equiv \pi\frac{\delta^2}{\delta^2 \Delta}
\int
\br{\prod_{i=1}^{n+1} \frac{d\vec q_i}{\pi \br{\vec q_i^2}^2}} \br{\prod_{i=1}^{n} \frac{d\beta_i}{\beta_i}}
I^A_{01} J_{12} J_{23} \cdots J_{n,n+1} I^B_{n+1,0},
\ea
where $\vec{\Delta}=\vec{q}_k-\vec{q}_{k+1}$.
Keeping these factors together we obtain:
\ba
m^4 \pi\frac{d\sigma^{\gamma\gamma}}{d\beta d^2\Delta}=\sum_1^\infty\frac{\alpha^{2(n+2)}}{4\pi(2\pi^2)^n}
F_{k,n}(\beta)G_{k,n}(x), \qquad x = \frac{\vec{\Delta}^2}{m^2}.
\ea
 The lowest order inclusive cross section
 of lepton pair production in $\gamma\gamma$ collisions for the case of unpolarized photon beams is:
 \ba
m^4 \beta \pi\frac{d\sigma_1^{\gamma\gamma}}{d\beta d^2\Delta}=\br{\frac{\alpha^2}{2\pi}}^3 G_1(x) L_e, \nn
\ea
\ba
G_1(x)&=&
\frac{16}{9} m^4\int\frac{d^2q}{\pi}\int\limits_0^1dx_1\int\limits_0^1dx_2\int\limits_0^1dx_3\int\limits_0^1dx_4
\times\nn \\
&\times&
\br{\frac{1+a_1}{m^2+a_1\vec{q}^2}}\br{\frac{1+a_2}{m^2+a_2 (\vec{q}+\vec{\Delta})^2}}
\br{\frac{a_3+a_4-5a_3a_4}{m^2+a_3\vec{q}^2+a_4(\vec{q}+\vec{\Delta})^2}},
\label{G1Definition}
 \ea
\ba
    && a_1 = x_1 \br{1-x_1}, \qquad a_3 = x_3 \br{1-x_3}, \nn \\
    && a_2 = x_2 \br{1-x_2}, \qquad a_4 = x_4 \br{1-x_4}, \nn
\ea
where $x=\Delta^2/m^2$.
Numerical value of the dimensionless quantity $G_1(x)$ is given in Fig.~\ref{FigOfG1}.
Numerical integration of this function over all possible tranverse momentum gives
\ba
\int\limits_0^\infty
dx G_1(x) \approx 26.5.
\ea

\section {Summation on the pair number}

Using representation of logarithmic enhancement factor $L_e^n/n!$
in the form of the integral over complex variable (see (\ref{OintRepresentation}))
the cross section of arbitrary number of pair production in $\gamma e$, $\gamma\gamma$ and $ee$
collisions may be written down in the universal form (\ref{CSwithArbitraryPairsNumber}),
where function $\Psi_j(x_1,x_2)$ is defined as:
\ba
    \Psi_j(x_1,x_2) &=&
    \sum_{n=1}^{\infty} h^{-n}
    \br{\prod_{i=1}^n \int \frac{d^2 q_i}{\pi \br{q_i^2}^2}}
    J_{12} \cdots J_{n-1,n}. \\
    h^{-1} &=& \frac{\alpha^2}{2\pi^2 j}.
\ea
This function are based on the integral equation:
\ba
    \Psi_j(x_1,x_2) =
    k(x_1,x_2) +
    h^{-1} \int\limits_0^\infty dx_1'
    k(x_1,x_1') \Psi_j(x_1',x_2).
\ea
There are two different contribution to kernel $k(x_1,x_2)$
(we do not consider here $\pi^0$ channels):
$k(x_1,x_2) = k_e(x_1,x_2) + k_\pi(x_1,x_2)$,
with
\ba
k_e(x_1,x_2) &=&
4
\int\limits_0^1 d\alpha
\int\limits_0^1 d\beta
\frac{a+b-5ab}{a x_1 + b x_2 + m^2}, \label{KernelKE}\\
k_\pi(x_1,x_2) &=&
2
\int\limits_0^1 d\alpha
\int\limits_0^1 d\beta
\frac{1/4 - (a+b-5ab)}{a x_1 + b x_2 + m^2}. \label{KernelKPi}
\ea
The function $\Psi_j(x_1,x_2)$ dependence of variable $j$
have a square root singularity in $j$-plane.
To find the most right position of this cut
in $j$-plane we may investigate the
homogenous integral equation:
\ba
    \Delta \Psi_j^i(x_1,x_2) =
    \frac{1}{\kappa^i} \int\limits_0^\infty dx_1'
    k^i(x_1,x_1') \Delta\Psi_j^i(x_1',x_2), \qquad
    i=e,\pi.
\ea
Further simplification consists in putting the mass of created particles equal to zero $m=0$.

To solve this homogeneous integral equation let expand function $\Psi_j(x_1,x_2)$ by full set of functions
$(x_1/x_2)^\lambda/\sqrt{x_1 x_2}$, with $\lambda$  pure
imaginary parameter and find the relevant eigenvalue..

First we can perform the integration on the intermediate photon momentum. We have:
\ba
\int\limits_0^\infty \frac{t^\gamma dt}{a+bt}=\frac{\pi}{\sin(\pi\gamma)}a^\gamma b^{-\gamma-1},
\qquad a, b > 0,
\ea
with $\gamma=-1/2+i\eta$. Using the explicit form of  kernels
$k_e$, $k_\pi$ (see (\ref{KernelKE}), (\ref{KernelKPi})) and
performing $\alpha$, $\beta$ integration after some algebra we obtain:
\ba
\kappa^e=\frac{\pi^2(11+12\eta^2)\sinh(\pi\eta)}{16\eta(1+\eta^2)\cosh^2(\pi\eta)}; \nn \\
\kappa^\pi=\frac{\pi^2(5+4\eta^2)\sinh(\pi\eta)}{32\eta(1+\eta^2)\cosh^2(\pi\eta)}.
\ea
As a result we obtain the following solution:
\ba
\left.\Psi_j(x_1,x_2)\right|_{m^2=0} =
\frac{1}{2\pi i}
\int\limits_{-i \infty}^{i \infty} d\lambda
\br{\frac{x_1}{x_2}}^\lambda \frac{1}{\sqrt{x_1 x_2}}
\br{\frac{1}{\kappa(\lambda)} - \frac{1}{h(j)}}^{-1},
\ea
where
\ba
\kappa(\lambda) = \frac{\pi^2 \sin(\pi \lambda)}{128 \lambda (1-\lambda^2) \cos^2(\pi\lambda)}
\brs{2 N_f \br{11-12\lambda^2} + N_b\br{5 - 4 \lambda^2}},
\ea
where $N_f$ is the number of charged leptons and $N_b$
is the number of charged mesons.
The solution if the analytical function of $j$ with the
cut located at $1 < j < j_0 = 1 + \frac{\pi \alpha^2}{64}\brs{22 N_f + 5 N_b}$.
This formula is the generalization of formula obtained in \cite{Frolov:1970ij, Gribov:1970ik, Lipatov:1971ax}
where only charged fermions case was considered.

\section {Conclusion}
\label{Conclusion}

The partially integrated cross sections, which takes into account the
transfer momentum not exceeding $\brm{\vec q_1^2} < Q_1 = \sqrt{-q_1^2}$,
$\brm{\vec q_2^2} < Q_2 = \sqrt{-q_2^2}$ are given in Appendix~\ref{AppendixTotalCrossSections}.
We choose $L_\pi \approx 20$ which is expected at LHC. It can be seen
that the cross sections are big enough to be taken into account and measured.

We should note that the deviation from this
asymptotic formulae can be considerable. For instance the
cross section of process $pp\to pp e \bar e$
(see formula (\ref{ForRacah}) in Appendix~\ref{AppendixTotalCrossSections})
gives $\sigma^{LL}_{pp\to pp e \bar e} = 9.9\mb$ at $\sqrt{s}=14~\mbox{TeV}$,
while the more precise form is \cite{Racah:1936}:
\ba
\sigma^{pp\to pp e \bar e} &=&
\frac{28}{27} \frac{\alpha^4}{\pi m_e^2}
\br{L_p^3 +A L_p^2 +B L_p +C },
\ea
where $L_p=\ln\br{s/M_p^2}$ and
\ba
A &=&
-\frac{178}{28} = -6.35714,
\nn\\
B &=&
\frac{1}{28}\br{7\pi^2 + 370} = 15.6817,
\nn\\
C &=&
-\frac{1}{28}\br{348 + \frac{13}{2}\pi^2 - 21 \zeta_3} = -13.8182,
\nn
\ea
which gives $\sigma_{pp\to pp e \bar e} = 7\mb$.
Again for process $e\bar e \to 2 e 2\bar e$ we have following
precise result \cite{Kuraev:1973sj}:
\ba
\sigma^{e\bar e \to 2 e 2\bar e} &=&
\frac{\alpha^4}{\pi m_e^2}
\br{A L_e^3 +B L_e^2 +C L_e +D },
\ea
where $L_e=\ln\br{s/m_e^2}$ and
\ba
A &=&
\frac{28}{27} = 1.03704,
\nn\\
B &=&
-\frac{178}{27} = -6.59259,
\nn\\
C &=&
-\frac{164}{9}\zeta_2 + \frac{490}{27} = -11.8262,
\nn\\
D &=&
\frac{52}{3}\zeta_2 \ln 2 + \frac{401}{9}\zeta_3 +
\frac{886}{27}\zeta_2 - \frac{428}{27} = 111.448.
\nn
\ea
In leading logarithmical approximation it gives
$\sigma^{LL}_{e\bar e \to 2 e 2\bar e} = 56\mb$, while
the precise value is
$\sigma_{e\bar e \to 2 e 2\bar e} = 45\mb$
at $\sqrt{s}=14~\mbox{TeV}$.

So the formulae given above have the accuracy of order $\pm (15-20) \%$.

\section {Acknowledgements}
Two of us (E.A.K. and Yu.M.B.) are grateful to INTAS grant (05-1000008-8328)
and Belorussian grant MK-2952.2006.2 for financial support.
We are also grateful to S.~Dubnicka, C.~Adamuscin,
A.~Z.~Dubnickova for useful discussions.
We also grateful to Lev Lipatov for critical comments.

\appendix

\section {Weizsaecker-Williams (WW) enhancement for $\gamma e$ and
$e^+e^-$ collisions}
\label{AppendixWW}

Let us consider the details of $e\to\gamma^* e$ vertex
\ba
e(p_1) \to \gamma^*(q) + e(p'_1) \nn
\ea
in peripherical kinematics. Using Sudakov parametrization of
the 4-momentum of virtual photon:
\ba
q = \alpha p + \beta \tilde p_1 + q_\perp,
\quad
p^2 = \tilde p_1^2 = 0,
\quad
q_\perp p = q_\perp \tilde p_1 = 0,
\quad
2p_1 \tilde p_1 = m^2,
\ea
and keeping in mind the electrons on mass shell conditions:
\ba
\br{p_1-q}^2 - m^2 = -s\alpha(1-\beta)-m^2\beta - \vec q^2 = 0,
\ea
we obtain for the square of virtual photon momentum:
\ba
q^2 = -\frac{1}{1-\beta} \brs{\vec q^2 + m^2\beta^2}.
\ea
We use here the standard definitions $q_\perp^2 = -\vec q^2$,
$s=2p p_1 \gg m^2, \vec q^2$. It is important to note that transfer momentum
is space-like vector with nonzero norm. Performing integration on
$d^2\vec q$ in matrix element square and keeping in mind the consequences of
gauge invariance we obtain an enhancement factor in rapidity $\beta$:
\ba
    \int\limits_0^{Q^2}
    \frac{\vec q^2 d\vec q^2}
    {\br{\vec q^2 + m^2 \beta^2}^2}
    \approx \ln \frac{Q^2}{m^2\beta^2}.
\ea

\section{The total cross sections of $\pi_0$ and $\pi_+\pi_-$ production}
\label{AppendixTotalCrossSections}

Let us consider the total cross sections of processes of pions production within
logarithmic accuracy. In case then the momentum transfer does not exceed the $Q_1, Q_2 < M_\pi$
we have (see comments to formula (\ref{Tot})):
\ba
\sigma^{ee\to ee \pi^0} &=&
\frac{\alpha^4 L_\pi}{12 \pi^3 f_\pi^2}
\brs{L_\pi^2 + 6 \ln\br{\frac{M_\pi}{m_e}} \ln\br{\frac{s}{M_\pi m_e}}},
\qquad Q_1 \sim Q_2 < M_\pi,
\nn\\
\sigma^{ee\to ee \pi^+\pi^-} &=&
\frac{4}{27}\frac{\alpha^4}{M_\pi^2}
L_\pi
\brs{L_\pi^2 + 6 \ln\br{\frac{M_\pi}{m_e}} \ln\br{\frac{s}{M_\pi m_e}}},
\qquad Q_1 \sim Q_2 < M_\pi,
\nn\\
\sigma^{pp\to pp \pi^0} &=&
\frac{\alpha^4 L_\pi^3}{12 \pi^3 f_\pi^2},
\qquad Q_1 \sim Q_2 < M_P,
\nn\\
\sigma^{ep\to ep \pi^0} &=&
\frac{\alpha^4 L_\pi^2}{12 \pi^3 f_\pi^2}
\brs{L_\pi + 3 \ln\br{\frac{M_\pi}{m_e}}},
\qquad Q_1 < M_\pi, \quad Q_2 < M_P,
\nn\\
\sigma^{\gamma p \to p 2 \pi^0} &=&
\frac{\alpha^5}{2^8 \pi^6 f_\pi^2}
\br{L_\pi^2+L_\pi \ln\br{\frac{M_\pi}{m_e}}} \int \frac{d^2 \vec q_1}{\pi f_\pi^2} R_{01}^4,
\qquad Q_2 < M_\pi,
\nn\\
\sigma^{\gamma e \to e \pi^0} &=&
\frac{\alpha^3}{4 f_\pi^2}
\ln\br{\frac{s}{M_\pi m_e}},
\qquad Q_2 < M_\pi,
\nn\\
\sigma^{\gamma e \to e \pi^+\pi^-} &=&
\frac{2\alpha^3}{9 M_\pi^2}
\ln\br{\frac{s}{M_\pi m_e}},
\qquad Q_2 < M_\pi,
\nn\\
\sigma^{\gamma e \to e 2\pi^0} &=&
\frac{\alpha^5}{2^8 \pi^6 f_\pi^2}
\br{L_\pi^2 + 2 L_\pi \ln\br{\frac{M_\pi}{m_e}}}
\int \frac{d^2 q_1}{\pi f_\pi^2} R_{01}^4,
\qquad Q_2 < M_\pi,
\nn\\
\sigma^{\gamma e \to e 2(\pi^+\pi^-)} &=&
\frac{\alpha^5}{\pi^2 M_\pi^2}
L_\pi
\ln\br{\frac{s}{M_\pi m_e}} J_\pi,
\nn\\
\sigma^{\gamma \gamma \to 2\pi^0} &=&
\frac{\alpha^4}{2^8 \pi^5 f_\pi^2}
\int \frac{d^2 \vec q_1}{\pi f_\pi^2} R_{01}^4,
\nn\\
\sigma^{\gamma \gamma \to 2(\pi^+\pi^-)} &=&
\frac{\alpha^4}{\pi M_\pi^2} J_\pi,
\nn\\
\sigma^{\gamma \gamma \to 3\pi^0} &=&
\frac{\alpha^6}{2^{13} \pi^9 f_\pi^2} L_\pi
\int \frac{d \vec q_1^2}{f_\pi^2}
\int \frac{d \vec q_2^2}{f_\pi^2}
R_{01}^2 R_{02}^2 R_{12}^2. \nn
\ea
The integrals
$R_{01}$, $R_{02}$, $R_{12}$ have a form:
\ba
R_{01} \br{\vec q_1^2} &=&
2M_q^2
\int\limits_0^1 dx_1
\int\limits_0^{1-x_1}
\frac{dx_2}{M_q^2 - M_\pi^2 x_2 x_3 + \vec q_1^2 x_1 x_3},
\qquad
x_3 = 1-x_1-x_2, \nn\\
R_{12} \br{\vec q_1^2,\vec q_2^2} &=&
2M_q^2
\int\limits_0^1 dx_1
\int\limits_0^{1-x_1}
\frac{dx_2}{M_q^2 - M_\pi^2 x_2 x_3 + \br{\vec q_1^2 x_2 + \vec q_2^2 x_3} x_1},
\ea
where $M_q$ is the loop quark mass ($M_q \sim 260\MeV$).
We keep into account only the logarithm of ratio
of pion and electron masses $\ln\br{M_\pi/m_e}$.
As rather good approximation (within 5\% accuracy) we can omit
term $\br{- M_\pi^2 x_2 x_3}$ compared with $M_q^2$ in
$R_{01}$ and $R_{12}$ denominators and obtain the following
formulae:
\ba
R_{01} \br{\vec q_1^2} &=&
\frac{M_q^2}{\vec q_1^2} \ln^2\br{\frac{b+1}{b-1}},
\qquad
b = \sqrt{1+\frac{4 M_q^2}{\vec q_1^2}}, \\
R_{12} \br{\vec q_1^2,\vec q_2^2} &=&
\frac{\vec q_1^2 R_{01}\br{\vec q_1^2} - \vec q_2^2 R_{01}\br{\vec q_2^2}}
{\vec q_1^2 - \vec q_2^2}.
\ea
The numerical evaluation of phase integrals gives:
\ba
\int \frac{d^2 \vec q_1}{\pi f_\pi^2} R_{01}^4
&\approx&
46.2 \nn\\
\int \frac{d^2 \vec q_1}{\pi f_\pi^2}
\int \frac{d^2 \vec q_2}{\pi f_\pi^2}
R_{01}^2 R_{02}^2 R_{12}^2
&\approx&
2521. \nn
\ea
The total cross sections in leading logarithmical approximation of processes
$pp\to pp e \bar e$ and $pp\to pp \pi^+ \pi^-$ have a form:
\ba
\sigma^{pp\to pp e \bar e} &=&
\frac{28}{27} \frac{\alpha^4}{\pi m_e^2}
\ln^3\br{\frac{s}{M_p^2}},
\label{ForRacah}
\\
\sigma^{pp\to pp \pi^+ \pi^-} &=&
\frac{4}{27} \frac{\alpha^4}{\pi M_\pi^2}
\ln^3\br{\frac{s}{M_p^2}}.
\nn
\ea
For completeness we evaluate this cross sections at energy $\sqrt{s} = 14\TeV$
(available at Tevatron). In that case
$L_e = 34$, $L_\pi = 23$ and the cross sections are:
\begin{tabular}{lclcl}
$\sigma^{ee\to ee \pi^0} = 12\nb$, & $\qquad$ &  $\sigma^{\gamma e \to e \pi^0} = 127\nb$, & $\qquad$ & $\sigma^{\gamma \gamma \to 2(\pi^+\pi^-)} = 2.47\pb$, \\
$\sigma^{ee\to ee \pi^+\pi^-} = 309\nb$, & $\qquad$ & $\sigma^{\gamma e \to e \pi^+\pi^-} = 53\nb$, & $\qquad$ & $\sigma^{\gamma \gamma \to 3\pi^0} = 1.64\ab$, \\
$\sigma^{pp\to pp \pi^0} = 4.28\nb$, & $\qquad$ & $\sigma^{\gamma e \to e 2\pi^0} = 0.14\pb$, & $\qquad$ & $\sigma^{pp\to pp e \bar e} = 9.9\mb$, \\
$\sigma^{ep\to ep \pi^0} = 7.38\nb$, & $\qquad$ & $\sigma^{\gamma e \to e 2(\pi^+\pi^-)} = 3.79\pb$, & $\qquad$ & $\sigma^{pp\to pp \pi^+ \pi^-} = 20\nb$, \\
$\sigma^{\gamma p \to p 2 \pi^0} = 0.117\pb$, & $\quad$ & $\sigma^{\gamma \gamma \to 2\pi^0} = 0.076\pb$. & \\
\end{tabular}

\section{Details of $J_{12}^e$ calculation}
\label{AppendixImpactFactor}

We put here some details of obtaining $J_{12}^e$ from formula (\ref{JE}).
After performing traces it can be written down in the form:
\ba
J_{12}^e &=& \int\limits_0^1 dx_+ \int \frac{d^2q_+}{\pi}
\left\{
    \frac{\br{D_1-D_2}^2 \br{\vec q_+^2+m^2}\br{\vec q_-^2+m^2}}{\br{D_1 D_2}^2} +
\right. \nn\\
&+& \frac{\vec q_1^2 \br{\vec q_+^2+m^2}}{D_1^2}
+ \frac{\vec q_1^2 \br{\vec q_-^2+m^2}}{D_2^2} +\nn\\
&+& 2 \br{\vec q_1 \vec q_-} \br{\vec q_+^2+m^2} \frac{D_1-D_2}{D_1^2 D_2}
- 2 \br{\vec q_1 \vec q_+} \br{\vec q_-^2+m^2} \frac{D_1-D_2}{D_1 D_2^2}
- \nn\\
&-&
\left.
 2 \frac{\vec q_1^2 \br{m^2 - \br{\vec q_+ \vec q_-}}+
2 \br{\vec q_+ \vec q_1}\br{\vec q_- \vec q_1}}{D_1 D_2}
\right\}.
\ea
Expression in figure brackets explicitly vanishes if $\vec q_1 \to 0$ or $\vec q_2 \to 0$.
Each separated term contain ultraviolet divergence and requires
regularization. That is the reason to enter the ultraviolet cut off
parameter $\Lambda$. For example the contribution of
second and third terms are:
\ba
&&\int \frac{d^2q_+}{\pi}
\br{
\frac{\vec q_1^2 \br{\vec q_+^2+m^2}}{D_1^2}
+ \frac{\vec q_1^2 \br{\vec q_-^2+m^2}}{D_2^2} } = \nn\\
&&
=\vec q_1^2
\brs{
2\ln\frac{\Lambda^2}{D_0} + \frac{\vec q_2^2 \br{x_+-x_-}^2}{D_0}
},
\ea
where $D_0 = x_+ x_- - \vec q_2^2 + m^2$.
In next terms we join the denominators using Feynman trick:
\ba
\frac{1}{D_1 D_2} = \int\limits_0^1 \frac{dt}{\br{t D_1 + \bar t D_2}^2} =
\int\limits_0^1 \frac{dt}{\br{\br{\vec q_+ - \vec b}^2 + D}^2},
\ea
where $\vec b = \bar t \vec q_1 + x_+ \vec q_2$,
and $D = m^2 + t \bar t \vec q_1^2 + x_+ x_- \vec q_2^2$.
The following integration on $\vec q_+$ is straightforward.
The dependence of ultraviolet cut off $\Lambda$ disappears.
The result of expression will contain terms with $\ln(D/m^2)$, $D^0$, $D^{-1}$, $D^{-2}$, $D^{-3}$.
Using the integration by parts
\ba
\int\limits_0^1 dt
\frac{\vec q_1^2 (1-2t) f(t)}{D^2}
 = -\left.\frac{f(t)}{D}\right|_0^1 +
 \int\limits_0^1 \frac{dt}{D} f'(t), \nn
\ea
we may express terms containing $D^{-2}$, $D^{-3}$
in terms with $D^{-1}$.
As a result we get:
\ba
J_{12}^e &=&
\int\limits_0^1 dx_+
\int\limits_0^1 dt
\left\{
2 b (1-4b) \br{\frac{1}{D} - \frac{1}{D_0}} \br{\vec q_2^2}^2
+\right.\nn\\
&+&\left.\frac{1}{D}
\brs{\vec q_1^2 \vec q_2^2 (1+8 a b - 2b) - 8a b \br{\vec q_1 \vec q_2}^2}
\right\},
\label{RawJ12e}
\ea
where $a = t \bar t$ and $b=x_+ x_-$.
The first term in figure brackets can be written in symmetrical form:
\ba
&&\int\limits_0^1 dx_+
\int\limits_0^1 dt
2 b (1-4b) \br{\frac{1}{D} - \frac{1}{D_0}} \br{\vec q_2^2}^2 =\nn\\
&&=
2 \vec q_2^2 \int\limits_0^1 dx_+
\br{x_+ x_- (1-2x_+)} \frac{d}{dx_+} \ln\frac{D}{D_0} =\nn\\
&&=
- \vec q_1^2 \vec q_2^2 \int\limits_0^1 \frac{dx_+ dt}{D}
\br{1-4a}\br{1-6b}.
\ea
Substituting this result to (\ref{RawJ12e}) we obtain (\ref{JE}).


\end{document}